\documentstyle[aas2pp4]{article}

\def\today{\number\day\enspace
     \ifcase\month\or January\or February\or March\or April\or May\or
     June\or July\or August\or September\or October\or
     November\or December\fi \enspace\number\year}
\def\clock{\count0=\time \divide\count0 by 60
    \count1=\count0 \multiply\count1 by -60 \advance\count1 by \time
    \number\count0:\ifnum\count1<10{0\number\count1}\else\number\count1\fi}

\lefthead{Barlow and Sargent}
\righthead{Keck Spectra of PKS 0123+257: Intrinsic Absorption}
 
\begin{document}

\title{Keck High Resolution Spectroscopy of PKS 0123+257:\\
       Intrinsic Absorption in a Radio Loud Quasar }

\author{Thomas A. Barlow\altaffilmark{1} 
    and W. L. W. Sargent\altaffilmark{1}}
\affil{Astronomy Department,
       California Institute of Technology,
       Pasadena, CA 91125}

\altaffiltext{1}{Visiting Astronomer, W.M.Keck Observatory,
                 jointly operated by the California Institute of Technology
                 and the University of California}

\begin{abstract}

We present results from Keck I high resolution spectroscopy of the
radio loud quasar PKS 0123+257 ($z_e$=2.364, V=17.5).  In this object
we detect Ly$\alpha$, \ion{N}{5}\  $\lambda\lambda$1238,1242,
\ion{Si}{4}\ $\lambda\lambda$1393,1402, and
\ion{C}{4}\ $\lambda\lambda$1548,1550 in an absorption system at a
redshift of 2.369.  The Ly$\alpha$ line has a square-bottomed profile
suggesting a high column density of gas, yet the line does not reach
zero intensity.  The resolved \ion{C}{4}\ doublet ratio also clearly
demonstrates that the absorbing clouds at this redshift do not fully
occult the background light source along our line-of-sight.

The absorption lines are positioned near the centers of the broad
emission-lines and the coverage fraction of the strongest absorption
lines varies inversely proportionally with the strength of the
corresponding emission lines.  This implies that although the
absorption-line region may obscure the continuum source, it does not
completely occult the broad emission-line region.  This effect
suggests that the lines are formed close to the QSO central region.
A model is proposed in which the {\it apparent} coverage fraction
derived for the weaker absorption lines may vary with the column
density of the lines.

Broad absorption-lines (which are known to be intrinsic) are found
nearly exclusively in radio-quiet objects.  Intrinsic narrow
absorption lines have previously been found in radio quiet QSOs; it
is therefore significant that an intrinsic absorption system has been
verified in a radio loud quasar.

\end{abstract}


\vspace{20pt}
\centerline{{\it Accepted to the Astronomical Journal (January 1997 issue.)}}

\section{Introduction}

PKS 0123+257 (V=17.5, $z_e$ = 2.364), or 4C 25.05, is a radio loud
quasar (\cite{sch68}; \cite{car76}.) This object shows a strong,
narrow absorption-line system with an absorption-line redshift
($z_a$) close to the emission-line redshift ($z_e$.)  We selected
this object for study with the High Resolution Echelle Spectrograph
(HIRES) on the Keck I telescope in an effort to investigate the
nature of intrinsic absorption systems.

We define an ``intrinsic'' absorption system as one which is caused
by clouds in the QSO environment, either near the continuum source or
elsewhere in the host galaxy.  This is to be distinguished from
systems which arise from intervening galaxies and intergalactic
clouds.  Intrinsic systems can either be very broad ($\sim$10,000 km
s$^{-1}$) as in the so called ``broad absorption-line'' (BAL) QSOs,
or they can be relatively narrow ($\sim$ 100 km s$^{-1}$) with a
number of cases being intermediate in width.

Intrinsic absorption systems can be used to study the QSO
environment, test standard models of quasars, and estimate
metallicities in quasar environments.  Recent work has strongly
suggested that gas near a QSO has heavy element abundances much
enhanced relative to the solar value (\cite{pet94}; \cite{ham96} and
references therein.) Intrinsic narrow lines are potentially more
useful in metallicity estimates than BALs since it is often possible
to separate the components of close doublets such as \ion{C}{4}\
$\lambda\lambda$1548,1550 and \ion{N}{5}\ $\lambda\lambda$1238,1242.
In practice it can be ambiguous as to whether a particular narrow
line absorption system with $z_a \sim z_e$ is intrinsic or
intervening.  For these reasons, it is important to develop means of
distinguishing the two categories.

The best candidates for intrinsic narrow lines are systems with $z_a \sim
z_e$ .  This category of ``associated'' absorption has been defined
as lines which appear within several thousand km s$^{-1}$ of the
emission redshift and are less than a few hundred km s$^{-1}$ wide
(cf. \cite{fol88}; \cite{and87}.) Since we expect some intervening
lines to appear close to the emission redshift, these ``associated''
lines may not always be intrinsic to the QSO.  Occasionally we
observe strong, complex absorption, commonly with high excitation,
which are thought to be intrinsic because such systems are rarely
seen at $z_a \ll z_e$ ({\it e.g.} PKS 1157+014, cf. \cite{bri84} and
GC 1556+335, cf. \cite{mor86}.) Previous work on narrow associated
absorption has attempted to establish the intrinsic nature of the
lines statistically.  There appears to be an excess of narrow lines
near $z_e$ in large samples of QSOs (\cite{wey79}; \cite{fol86}.)
Although this result has been questioned by other surveys
(\cite{sar88}), it is now apparent that systems with large total
equivalent width tend to be close to the emission redshift (cf.
\cite{fol88}; \cite{bar88}).  Furthermore, there appears to be an
excess of associated lines in radio-loud quasars (\cite{and87})
relative to radio-quiet QSOs; such an excess could only be explained
by an intrinsic origin.

The following ways have been proposed
to distinguish intrinsic from intervening absorption systems
(in order of decreasing reliability):  
(1) time variability (\cite{ham95}; \cite{hamet96b}),
(2) high electron density as derived from fine structure
lines as a distance indicator when combined with an ionization level
estimate (cf. \cite{bah68}; \cite{wil75}; \cite{mor86}), 
(3) partial coverage of the
background light source along our line-of-sight
by the absorption-line region
(cf. \cite{wam93};  \cite{wam95}; \cite{hamet96}; \cite{hamet96b}),
(4) spectropolarimetry revealing increased fractional
polarization in the lines relative to the continuum (so far only
observed in BALs, cf. \cite{goo95}, \cite{coh95}),
(5) velocity structure, {\it i.e.} breadth, complexity, and
``correlated'' or ``smooth'' profile shape across many components,
(6) higher ionization level, and
(7) higher metallicities ({\it e.g.} \cite{pet94}.)

Although time variability has been observed often in BALQSOs
(\cite{bar92}; \cite{bar93}), method (1) generally requires extensive
monitoring of a large sample of objects over at least a few years.
Method (2) requires the presence of fine structure lines (usually) of
\ion{C}{2} or \ion{Si}{2}.  Unfortunately, the high ionization and/or
low column density of many systems makes these lines difficult to
detect.  Method (4) requires long exposure times with low resolution
and can only be used on objects with relatively high polarization
and, therefore, has only been useful for studying a few wide BALs.
Method (6) and (7) are easy to apply but are very sensitive to
photoionization models ({\it e.g.} \cite{ham96}) and there may be a
large overlap in these qualities between intrinsic and intervening
systems.  Fortunately, methods (3) and (5) can be applied directly
and easily using Keck and HIRES.  Recently it has become evident that
high resolution spectra with high signal-to-noise ratios can be a
useful tool in distinguishing intrinsic narrow line systems
(\cite{bar95}; \cite{hamet96}.) In the case of some narrow lines with
$z_a \sim z_e$ the lines appear to be smoother with less component
structure than most intervening systems; they also exhibit an effect
which implies that the absorption-line region does not fully occult
the background light source.  This effect is inferred from the
profiles of resolved lines in doublets such as \ion{C}{4}\ and
\ion{N}{5}.  Motivated by these considerations we investigate here
the nature of the associated lines in the radio-loud quasar PKS
0123+257.

\section{Observations}

The HIRES spectra for PKS 0123+257 were obtained during service
observing with the Keck I telescope on September 11 and 12, 1995.  We
acquired 4 exposures of 40 minutes (in cloudy skies and a full moon).
Three exposures were obtained with one echelle grating angle and 
one exposure with a different echelle angle in order to get complete
wavelength coverage of the region 3700 to 6100 \AA .

The spectra were detected as 38 echelle orders on a 2048$^2$
Tectronics CCD.  Each exposure used a 0.86 arc-second slit (3 pixels)
for a resolution of approximately 6.3 km s$^{-1}$.  The actual
resolution varies from about 6.0 to 6.5 km s$^{-1}$ FWHM due to the
change in dispersion across each echelle order.  The signal-to-noise
ratio (S/N) per 3 pixel resolution element ranged from 9 at
3900 \AA\ to 26 at 5400 \AA .  The S/N also varies with the distance
{}from the blaze on each order and the overlap between two setting
angles.  In particular, the setting with a single exposure had
significantly lower S/N which creates low S/N regions in the red
portion of the spectrum.

The data were reduced using our own reduction package which
automatically traces the orders, removes cosmic ray events, and uses
an optimal weighting scheme similar to that of \cite{hor86} for
extracting the spectra (cf. Barlow 1996, in preparation.)  The
wavelength calibration used low-order polynomial fits to a ThAr arc
lamp exposure in each order, loosely constrained by the fits in
adjacent orders.  The calibration error is about 0.1 pixel (0.2 km
s$^{-1}$) root-mean-square with up to a maximum error of about 0.3
pixels (0.6 km s$^{-1}$.)  This is the relative error across the
spectrum.  The absolute wavelength error is about 0.3 pixels (0.6 km
s$^{-1}$),  estimated from the night sky emission lines.  Altogether
there can be up to $\sim$1 km s$^{-1}$ error in the absolute
wavelength value at any given pixel.  The wavelengths are shifted to
a vacuum, heliocentric scale.

Due to imperfect corrections for variations in the pixel-to-pixel
response (flat field division), we estimate a error in the relative
level of the spectrum of about 0.5\% which appears to cause
``undulations'' on the scale of several pixels.  This is estimated
{}from spectra of bright stars taken previously with HIRES.  In
addition, there are fluctuations due to our technique of forcing
adjacent echelle orders to match over common wavelengths.  It is
estimated that systematic flux errors of $\sim$10\% may be present
over $\sim$1000 km s$^{-1}$ regions where the orders have overlapped
in wavelength.

Figure 1 shows the HIRES data for PKS 0123+257 from the Ly$\alpha$
(\ion{H}{1} $\lambda$1215) broad emission-line (BEL) to the
\ion{C}{4}\  $\lambda$1549 BEL.  The data were not flux calibrated
and the spectrum has been continuum normalized.  Note the low S/N
regions in the red portion of the spectrum.  Despite the artificial
fluctuations left over from the order-splicing procedure, the
relative strengths of the Ly$\alpha$, \ion{N}{5}\ $\lambda$1240, and
\ion{C}{4}\ $\lambda$1549 BELs are evident.  The $z_a \gtrsim z_e$
absorption can be seen just redward of the most prominent emission
peaks.  An emission line redshift of $z_e=2.364$ is estimated by
averaging the values measured from the Ly$\alpha$\ and
\ion{C}{4}\ emission peaks.

\placefigure{fig1}

\section{Analysis}

\subsection{Incomplete Occultation}

In figure 2, we compare the various lines in the associated system on
a velocity scale in which zero velocity is defined at $z_e$ =2.364.
There are three main absorption lines with the strongest absorption
appearing at $z_a$=2.3693 or about 480 km s$^{-1}$ redward of $z_e$.
Note that the strongest Ly$\alpha$\ absorption line appears to have a
flat-bottomed structure suggesting that the line is saturated, yet it
does not reach zero intensity.  This is a indication that the
absorbing gas clouds do not completely cover the background light
source along our line-of-sight.

\placefigure{fig2}

The velocity shift of 600 km/s redward of the Ly$\alpha$
and \ion{C}{4} emission peaks does not necessarily 
imply infall toward the QSO.
It has been shown that the \ion{C}{4}\ and Ly$\alpha$\ BEL
peaks are systematically blue-shifted relative to the true redshift
of the QSO (as determined by narrow forbidden lines) by ~500 $\pm$200
km s$^{-1}$ (\cite{tyt92}.)  This implies that all the absorption
shown in figure 2 may be produced by material ejected outward from
the central QSO engine.  From other studies of line profile and
velocity shift correlations, we know that \ion{C}{4}\ can be shifted
by as much as 1000 km s$^{-1}$ or more.  However, the width of 3500
km s$^{-1}$ for the \ion{C}{4}\  BEL in PKS 0123+257 suggests that the
shift is closer to a few hundred km s$^{-1}$ (cf. \cite{bro94a}.)
This suggests that the strongest absorption line may be close to
zero outflow velocity with respect to the QSO.

If the absorbing gas temperature is $\sim$10,000 K, then
the thermal widths for Ly$\alpha$\ and \ion{C}{4}\  are about 21 and
6 km s$^{-1}$, respectively.  Thus, the main \ion{C}{4}\ feature
contains many thermal widths and is essentially resolved in our
spectra.  The main line is not smooth like conventional broad
absorption-lines and shows some indication for narrow components.
However, these components are not as distinct and well separated as
the components which are often seen in intervening absorption
complexes.  Overlapping components and apparently large line widths
($\sim$50 km s$^{-1}$) may be indicative of intrinsic rather than
intervening absorption (\cite{bar95}.)

The apparent partial coverage of the background light source can be
investigated using two or more resolved lines from the same ion.  The
optical depth ratios should be equal to the ratios of the oscillator
strengths of the lines (assuming $\lambda_1\simeq\lambda_2$) if the
absorption clouds cover the light source.  If we have resolved the
\ion{C}{4}\ lines we can calculate the optical depth by taking the
natural logarithm of the residual intensity ($R.I.$) with the
continuum and emission-line flux normalized to unity, $\tau=-{\rm
ln}(R.I.)$.  We demonstrate that this is {\it not} the case in the
PKS 0123+257 system.  We have scaled the optical depth of the
\ion{C}{4} $\lambda$1550 line by a factor of two after measuring the
residual intensity assuming complete coverage by the absorbing
clouds.  In figure 3, we have plotted the scaled $\lambda$1550 line
(thick line) over the unscaled $\lambda$1548 line.  This figure shows
that the 1548 line does not go deep enough to match the strength
predicted by increasing the optical depth of the red line by two.
Therefore, both the main absorption system (at 480 km s$^{-1}$) and
the bluer system (at 190 km s$^{-1}$) exhibit partial coverage.  The
error in the spectrum at the position of the \ion{C}{4} $\lambda$1548 line
at 330 km s$^{-1}$ is much larger than for the adjacent
lines (these wavelengths fell within the inter-order gap of the best
exposures).  Given the depth of the red component, the data is
consistent with a coverage fraction (see below) between 0.4 and 1.0.

Figure 4 shows the same analysis for \ion{N}{5}.  Although the data
at these wavelengths have somewhat lower S/N, the lines appear to be
consistent with complete coverage of the background light source.  As
a check of this analysis technique, we do the same comparison for the
\ion{C}{4}\ lines in an absorption system at $z_a$=2.301 which is
assumed to be due to an intervening absorber.  As shown in figure 5,
this absorption system is consistent with complete coverage.

\placefigure{fig3}

\placefigure{fig4}

\placefigure{fig5}

\subsection{Defining Fractional Coverage}

We can also calculate the coverage fraction ($C_f$) from the measured
residual intensities at the lowest points in the lines.  $C_f$ is
defined as the fraction of the background light source, in this case
the QSO continuum source or the emission-line region, apparently
covered by the absorbing clouds (as defined by the formulas given
below.) Therefore, the amount of light (on a normalized scale)
apparently leaking through the absorption-line region (ALR) would be
$1 - C_f$.  If the coverage fraction at wavelength $\lambda$ is
$C_f(\lambda)$, then the {\it true} optical depth of a line at
$\lambda$ with the observed residual intensity $R(\lambda)$ is given
by:
$$
\tau(\lambda) = -{\rm ln} \left( { { R(\lambda) - 1 + C_f(\lambda) }
                          \over{ C_f(\lambda)} } \right).
$$
We neglect any smoothing of the line caused by the finite resolution of
the spectrograph.
An {\it apparent} column density per unit velocity can be derived
as described in Savage and Sembach (1991):
$ N(v) \propto \tau / f \lambda $.  It is assumed that $f_{blue}/f_{red}$ = 2
and $\lambda_{blue} \simeq \lambda_{red}$.
Let $R_b$ be the residual intensity of the blue (stronger) line 
of the doublet and $R_r$ be the residual intensity of the red line.
We assume that $C_f$ is the same for both lines.
Equating
the column densities derived from each line, $N_{blue}(v)=N_{red}(v)$,
we have:
$$ 
      -{\rm ln} \left( { { R_b - 1 + C_f } \over{ C_f } } \right)
  = -2 {\rm ln} \left( { { R_r - 1 + C_f } \over{ C_f } } \right)  ,
$$
$$
C_f = { { 1 + {R_r}^2 - 2 R_r } \over{ 1 + R_b - 2 R_r } } .
$$
If $R_r\leq R_b$, we assume $C_f=1-R_r$ 
and if $R_b < R_r^2$, we assume $C_f$=1.

The measured values of $C_f$ for \ion{C}{4} are
0.94{\thinspace}$^{0.95}_{0.93}$ (1$\sigma$ error limits) at 480 km
s$^{-1}$ and 0.63{\thinspace}$^{0.65}_{0.62}$ at 190 km s$^{-1}$.  In
figure 3, we have marked (3$\sigma$ error limits) the value of
$1-C_f$.  In the case of the line at 480 km s$^{-1}$, the fact that
the mark is close to the bottom of the line indicates that the lines
have high optical depth ({\it i.e.} are nearly saturated.) The width
of the mark indicates the portion of the spectrum where the residual
intensity was measured.  The line at 190 km s$^{-1}$ also appears to
be nearly saturated even though the lowest part of the line only
reaches a residual intensity of about 0.3.

The measured values of $C_f$ for \ion{N}{5} are
0.98{\thinspace}$^{1.00}_{0.96}$ (1$\sigma$ error limits) at 480 km
s$^{-1}$ and 0.79{\thinspace}$^{1.00}_{0.62}$ at 190 km s$^{-1}$.  In
figure 4, we show the values for the \ion{N}{5}\ lines.  Here the
1$\sigma$ error limits in the $1-C_f$ value are shown.  Both systems
appear to be consistent with complete coverage.  Although the S/N is
much lower near the \ion{N}{5}\ lines, the measured value of $1-C_f$
is significantly lower ($\sim$2$\sigma$) than $1-C_f$ for
\ion{C}{4}\ for the strongest absorption line at 480 km s$^{-1}$.
The value of $C_f$ measured for the \ion{C}{4}\ line at 480 km
s$^{-1}$ is 0.94 (using the data between 472 and 492 km s$^{-1}$.)
For the bluer line at 190 km s$^{-1}$ (calculated between 182 and 196
km s$^{-1}$), $C_f$ is about 0.63.  Both of these values are
significantly less than one ($>3\sigma$.) For \ion{N}{5}\ $C_f$ is
measured to be 0.98, and is consistent (within 1$\sigma$) with
complete coverage.

For Ly$\alpha$ we do not have another Lyman line for comparison.
However, the flatness of the bottom of the line at 480 km s$^{-1}$
suggests that it is saturated.  Using the bottom of the line as an
estimate of the fractional coverage we get $C_f$=0.91.  (This is only
a lower limit if the line is not saturated.)  This value is slightly
less than for \ion{C}{4}.  For \ion{Si}{4}\ $C_f$ is
0.93{\thinspace}$^{1.00}_{0.88}$ (with 1$\sigma$ errors).  The noise
near this line and the fact that the Silicon lines may not be as well
resolved as the Carbon lines (assuming that a significant portion of
the width is due to thermal broadening, rather than just cloud
turbulence) makes this $C_f$ value rather uncertain.

In table 1, we show our measured values for the residual intensities
at the lowest point in the lines and our estimates for the coverage
fractions.  We also calculate column densities assuming $C_f$=1.
Since we know that the coverage may be incomplete for all these
lines, these values are actually lower limits.  In column 7, we
present conservative lower limit estimates using the values of
$C_f$.  The fact that the \ion{C}{4}\  blue line yields a lower value
than the red line indicates that the lines are saturated.  In this
case, we would need to fit the lines with Voigt profiles to get
accurate estimates of the column densities.  Unfortunately, it is
difficult to determine the component positions and FWHM values for
such a complex system with several overlapping components.  However,
we do know that log(N(\ion{H}{1}))$\lesssim$18.5 because a
higher column density would show damping wings which are inconsistent
with the data.

\placetable{tbl-1}

We make the hypothesis here that the filling-in or partial coverage
of the lines is due to light from the broad emission-line region
(BELR) leaking through (or around) the ALR.  From the depths of the
lines it is clear that at least some of the BELR is occulted by the
absorbing clouds (i.e. the residual intensity at the bottom of the
absorption lines is much less than the strength of the BELs.)  In
this scenario it is reasonable to assume that all of the continuum
source is covered since the continuum producing region is generally
thought to be much smaller than the BELR (by a factor on the order of
100, cf.  \cite{ost93}.) However it is expected that this simple
assumption will not be strictly valid, since multiple regions of
varying dimensions may contribute to the BEL and the continuum
emitting regions (cf. \cite{wil93}.) This hypothesis can be tested in
that $C_f$ we calculate should be related to the strength of the
BEL at the position of the absorption line.  This is qualitatively
consistent with the fact that $C_f$ is nearly unity for
\ion{N}{5}\ (weak BEL), larger for \ion{C}{4}\ (strong BEL), and
larger still for Ly$\alpha$ which shows the highest emission line
peak.

\subsection{Incomplete Coverage of the Broad Emis- sion-Line Region}

Let us assume a relatively simple model where the amount of light at
the wavelength of the absorption line which is leaking through or
around the ALR is proportional to the strength of the BEL at that
wavelength.  At the position of the \ion{C}{4}\ absorption, the BEL
level is 2.0 in continuum normalized units and $C_f$ is 0.94 (where
the BEL plus continuum has been normalized to one.) This means that $
( ( 1 - 0.94 ) \times 2.0 ) / ( 2.0 - 1 ) = 0.12 $ of the BEL light
is not occulted by the absorption clouds.  For \ion{N}{5}\ the BEL
level is 1.3.  This means that we would expect $C_f = 1 - ( (0.3
\times 0.12) / 1.3 ) = 0.97$, which is consistent with our results.
For Ly$\alpha$ the BEL is 2.9 from which we expect a $C_f = 1 - ( (
1.9 \times 0.12 ) / 2.9 ) = 0.92 $, consistent with our estimated
value of 0.91.  However, this simple model does not apply to all the
$z_a \sim z_e$ absorption lines.  If we consider the line at 190 km
s$^{-1}$ redward of the emission redshift (see figure 3), we find
that the \ion{C}{4}\ line has $C_f$=0.63.  This means that $ ( 1 -
0.63 ) \times 2.0 ) / ( 2.0 - 1 ) =  0.74$ of the BEL light leaks
through (assuming it covers the continuum completely).  The
\ion{N}{5}\ line is too uncertain for comparison, but it is evident
that the Ly$\alpha$ is much deeper than predicted.  The predicted
value is $C_f = 1 - ( ( 1.9 \times 0.74 ) / 2.9 ) = 0.48$, but the
actual $C_f$ for Ly$\alpha$ must be at least 0.85 (see figure 2.)

We note here a few factors
which may cause deviations from the simplest interpretation.  
(1)
The position of the absorption lines over the BEL will effect our
calculations.  For example, since the emission is greater at the
position of $\lambda$1548 than $\lambda$1550, 
this means that $C_f$ values may be
slightly over-estimated.  In the case of \ion{Si}{4} the opposite is true,
since the emission is weaker at the position of the blue line of the
doublet.  This is presumably due to the \ion{O}{4}] 1402 BEL and means that
$C_f$ may be slightly larger than calculated.  And, in the case of
\ion{N}{5} , we have a BEL contribution from the red wing of Ly$\alpha$.
(2)
A complex geometry for the BELR would imply that the size of the BELR
(and hence the ability for the ALR to cover it) may
vary with velocity and ion.  In particular, there may be a narrow and broad
component to the \ion{C}{4}\ BEL such that the broad component is blue
shifted and arises in an emitting region closer to the QSO central
engine than the narrow component (\cite{bro94b}.)
(3)
The ``effective'' size of the ALR may vary with
ionization level and/or optical depth.  
For example, the geometry of the portion of the
region with relatively high amounts of \ion{Si}{4}\ may be different than
the geometry of the portion of the ALR with relatively
high amounts of \ion{C}{4}.  In another study of narrow intrinsic
lines in the QSO UM 675 (\cite{hamet96}), it is
noted that $C_f$ for \ion{C}{4}\ must be substantially less than that
for Ly$\alpha$.

We can reconcile the earlier results by appealing to factor (3).  We
hypothesize that the ALR has a fixed boundary which is smaller than
the apparent size of the BELR as seen along our line-of-sight.  This
means that there is a maximum line depth which occurs when the
optical depth is very high and the clouds occult all lines-of-sight
through the ALR.  This is the case for the strongest absorption lines
in PKS 0123+257, {\it e.g.} \ion{C}{4}, \ion{N}{5}, and Ly$\alpha$.
However, if the column density of absorbing gas varies between
different lines-of-sight through the ALR, then at lower optical depth
we get an {\it effective} $C_f$ which is significantly smaller.  This
is because the various lines-of-sight will contribute much
differently to the line profile.  In this scenario, we imagine that
the observed absorption line is composed of a superposition of
profiles from clouds with a range of column densities and a range of
$C_f$.  For example, we suppose that there exists two sub-regions in
the ALR which cover different lines-of-sight.  If we could look only
through sub-region \#1 we would measure an optical depth of $\tau\sim
2$, and if we looked only through sub-region \#2 we would measure
$\tau\sim 0.1$.  In practice, we observe the sum of the light going
through both sub-regions. If the first sub-region covers only 50\% of
the source then would measure $C_f \sim 0.5$, since the second
sub-region contributes little to the observed line profile.  Let us
now increase the optical depth of both regions by a factor of 5.
Since the optical depth in sub-region \#1 was already fairly high,
these clouds contribute only a little more to the depth of the line,
but the clouds of the sub-region \#2 now contribute much more.  The
line becomes deeper and $C_f$ increases.  This means that the {\it
effective} $C_f$ can vary depending on the optical depth of a given
transition without changing the geometry of the clouds or the
region.  In this scenario, factor (3) can cause a variation in the
effective $C_f$ for different ions {\it and} for different lines of
the same ion.  Note that this complicates what we mean by partial
coverage.  In this model, we cannot define partial coverage as simply
the fraction of sight lines which pass through absorbing gas.

Applying these ideas to our results, the system at 190 km s$^{-1}$
(see figure 2) has clouds which can cover at least 85\% of the
background light source as shown by the Ly$\alpha$ line.  However,
the lines-of-sight through this region have varying column density
such that the less abundant C$^{+3}$ ions only contribute
significantly to the line strength in certain sub-regions.  These
sub-regions are only able to effectively cover about 63\% of the
background light source.  Note that in this model we do {\it not}
need to vary the ionization level among the clouds in the ALR, we
only need to vary the column density looking through different
lines-of-sight toward the light source.

We can also use this model to explain the \ion{Si}{4}\ in the main
system at 480 km s$^{-1}$.  Above we measured $C_f$ of
\ion{Si}{4}\ to be 0.93 with a large error.  This value appears to be
larger than the value for \ion{N}{5}\ even though the \ion{N}{5}\ BEL
is stronger than the \ion{Si}{4}\ BEL.  This is inconsistent with the
simple analysis we discussed at the beginning of this subsection for
this absorption system.  The explanation here is that the line is
weak and thus we must appeal to factor (3), {\it i.e.} the Si$^{+3}$
ions in the lower column density lines-of-sight to not contribute
substantially to the line depth, and therefore for \ion{Si}{4}\ we do
not get a $C_f$ which is proportional to the BEL strength.

\section{\bf Discussion}

\subsection{Difficulty of Determining Column Densities}

Once we have absorbing clouds which do not cover the background light
source a range of effects must be considered when interpreting the
depth and profile of a line for the purpose of deriving column
densities.  When all the effects discussed above are considered, we
can appreciate the difficulty in determining accurate column
densities.  If we were to calculate the column densities assuming
complete coverage in this object, we would get values for \ion{C}{4},
\ion{N}{5}\ and Ly$\alpha$\ from which we could estimate ionization
levels and abundances.  In reality these values are inaccurate since
the lines are actually nearly saturated and the varying depths are
not caused simply by differences in ionic abundance, but rather they
are determined largely by differences in the fractional coverage of
the BELR.

Recently there has been work done on the metallicity and ionization
levels in intrinsic absorption systems.  In 0226-1024, Korista et
al.\ (1996) use ground-based and HST spectra of numerous transitions
to deduce enhanced metal abundances for the absorbing gas relative to
solar.  In UM 675, Hamann et al.\ (1995) argue for stratified regions
of greatly differing ionization levels.  In PG 0946+301,  Junkkarinen
et al.\ (1995) argue for Phosphorus abundances two orders of
magnitude greater than solar.  And using high resolution spectroscopy
of two QSOs, Petitjean et al.\ (1994) argue for increased
metallicities for absorption systems near the emission-line
redshift.  The success of line fitting and ionization estimates for
these systems indicates that these results are, in general,
accurate.  However, all of these results should be considered in the
context of the problems with studying intrinsic absorption lines as
discussed above.  The possibility of partial coverage has been
discussed in some of these works, but not yet applied.

\subsection{Determining the Causes of Partial Covering}

We discussed above the nature of the coverage fraction varying among
the lines.  This effect can result from a number of causes.  If, for
example, we were to see a different $C_f$ for \ion{Si}{4} and
\ion{C}{4}, we could explain this result if the ionization or the
optical depth varies between different lines-of-sight through the
ALR.  It will be possible to test which of these effects is dominant
by studying $C_f$ deduced from lines of different ions and different
lines due to the same ion.  We note that if this model is valid, then
even the two members of the \ion{C}{4}\ doublet will have different
$C_f$, making the determination of $C_f$ and the column densities
even more problematic.  Furthermore, one would expect $C_f$ to also
change in a time variable system if the variability is due to a
change in ionization level, {\it i.e.} as the fractional abundance of
an ion varies so does the optical depth in the line.  This effect can
be tested using high resolution spectra of a time variable absorption
system (as seen in the QSO 2343+1232, \cite{hamet96b}.)

\subsection{Failure to Distinguish Intrinsic Systems}

None of the seven methods mentioned in the introduction are
conclusive tests for intrinsic vs. intervening systems.  For example,
in the QSO 2116-358, \cite{wam93} show a system which suggests
partial coverage of the BEL in the \ion{Si}{3}\ $\lambda$1206 line;
yet the same system does not show absorption from excited fine
structure lines and thus must be a large distance ($>$1 kpc) from the
QSO nucleus.  It will be necessarily to investigate these seven
qualities statistically in many objects to determine how best to
discriminate intrinsic absorbers.

\subsection{Radio-Loud Quasars}

An excess of $z_a\sim z_e$ systems in radio-loud quasars relative to
radio-quiet quasars has been noted by previous studies (cf.
\cite{and87}).  In this paper, we have shown that a $z_a\sim z_e$
system in a radio-loud quasar shows a property (namely $C_f<1$) in
common with other intrinsic systems seen in radio-quiet quasars.
While there is an excess of $z_a\sim z_e$ systems in radio-loud
quasars, there is a definite lack of $z_a \ll z_e$ systems ({\it
i.e.} BALs.)

Recent work on high redshift radio galaxies has shown a frequent
occurrence $z_a\sim z_e$ Lyman-$\alpha$ absorption (\cite{oji96}).
The large fraction of galaxies with such absorption (60\%) implies
that the absorbing gas is intrinsic to the host galaxy.  Long-slit
spectra show that the absorbing region extends over the entire
emitting region (up to $\sim$50 kpc.)  Also, a correlation is found
between the size of the radio jet and the Lyman-$\alpha$ emitting
region.  What is the connection between the radio jet and the gas
giving rise to associated absorption?  Why is there a lack of
accelerated ($z_a \ll z_e$) intrinsic systems in radio bright
sources?  How are the emitting and absorbing regions related in
quasars?  We may be able to answer these questions by identifying and
studying intrinsic systems in both radio-loud and radio-quiet
objects.

\section{Summary}

In this paper we have presented high resolution spectroscopy of the
$z_a \sim z_e$ intrinsic absorption system in the radio quasar PKS
0123+057.  The data are consistent with the conclusion that the
background light source (which in this case includes both the
continuum and broad emission-line region) is not fully occulted by
the gaseous region causing this absorption.  This is deduced both
{}from the profile of the Ly-$\alpha$ absorption line and the residual
intensities of the \ion{C}{4} doublet.  The derived coverage
fractions, $C_f$, of the strongest absorption lines are inversely
proportional with the strength of the corresponding emission-lines,
which indicates that the absorption-line region obscures no more than
about 88\% of the light from the emission-line region.  Furthermore,
the coverage fractions measured for the weaker lines are consistent
with a model where the absorption-line region is inhomogeneous through
the various lines-of-sight towards the emission-line region which
pass through the absorbing gas.  This model shows that, for weak
lines, the calculated value of $C_f$ decreases with the depth of the
absorption line.  These effects mean that the optical depths measured
using the assumption that the absorbing gas homogeneously and
completely covers the background light source would significantly
underestimate the true column density through the absorption-line
region.  Furthermore, the degree to which the column densities are
underestimated will vary with the strength of the absorption line.
Accurate analyses of metal abundances in the gaseous environment near
QSOs using intrinsic absorption systems must necessarily account for
these factors.

\acknowledgements

We thank T. Bida and R. Campbell of Keck Observatory for obtaining
these data via service observing.  We are grateful to F.  Hamann, V.
Junkkarinen, and R. Cohen for useful discussions.  We also thank N.
Arav and D. Hogg for discussions regarding partial coverage of
intrinsic absorption lines.  This work was supported by grants
AST92-21365 and AST95-29073 from the National Science Foundation.

\clearpage

\figcaption[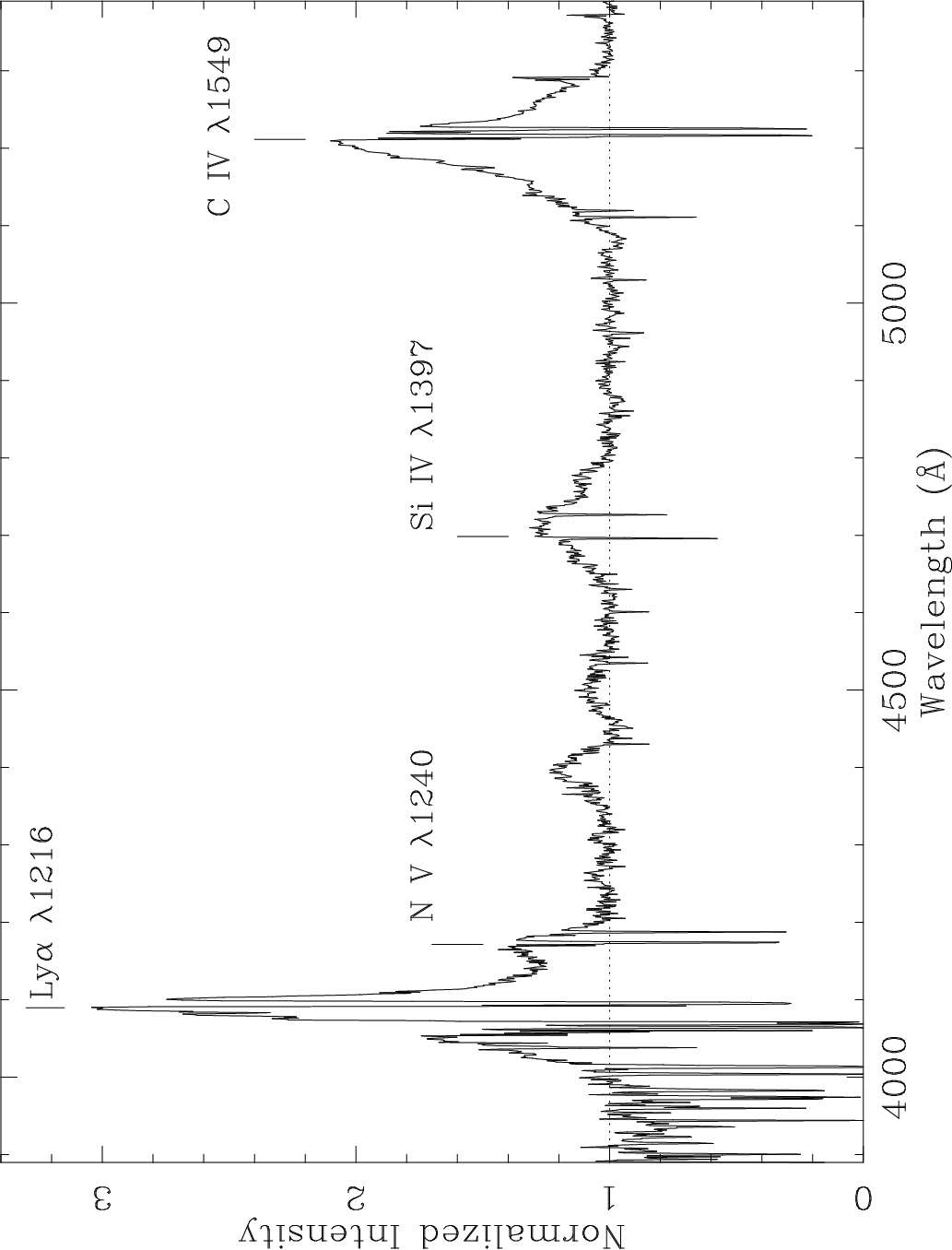]{HIRES spectrum of PKS 0123+257 with the
prominent emission lines marked.  The data has been binned such that
each pixel is 21 km s$^{-1}$ wide.  The continuum (but not the
emission-lines) has been normalized to one.  The associated
absorption lines can be seen just redward of the emission peaks.
\label{fig1}}

\figcaption[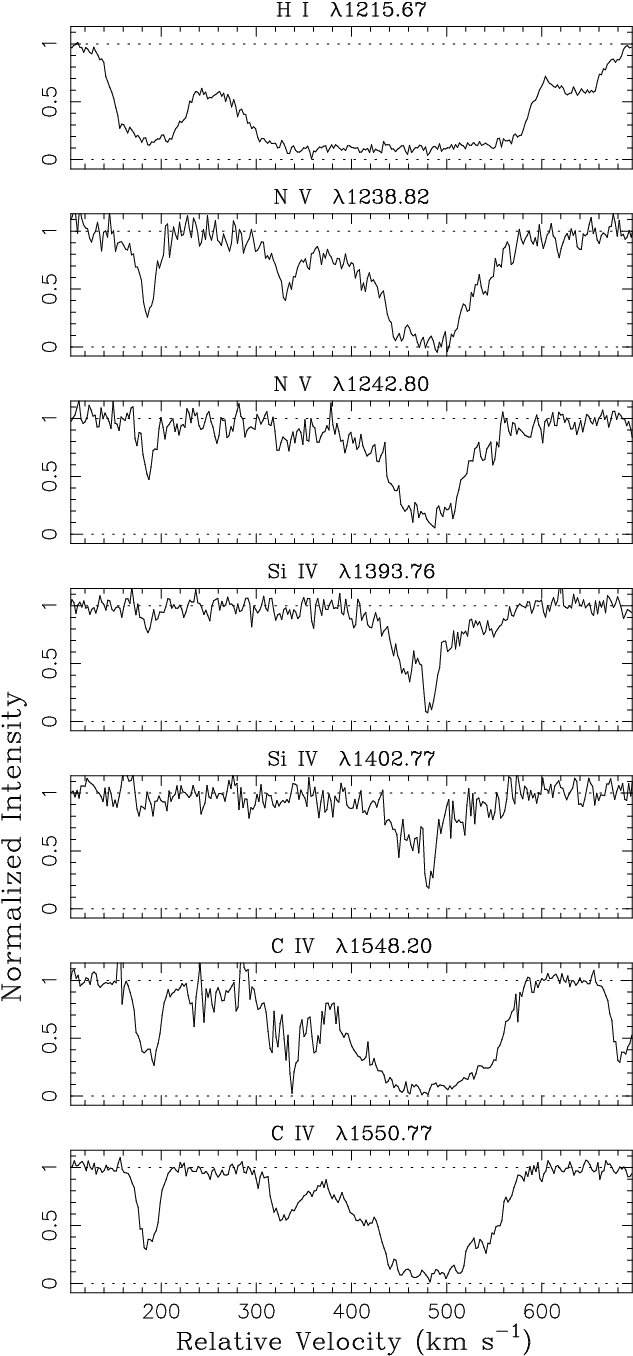]{The associated absorption system at $z_a$=2.369
shown on a velocity scale where zero velocity is at $z_e$=2.364.  The
continuum and emission line flux have been normalized to one.  There
are 3 pixels for each 6.3 km s$^{-1}$ FWHM resolution element. \label{fig2}}

\figcaption[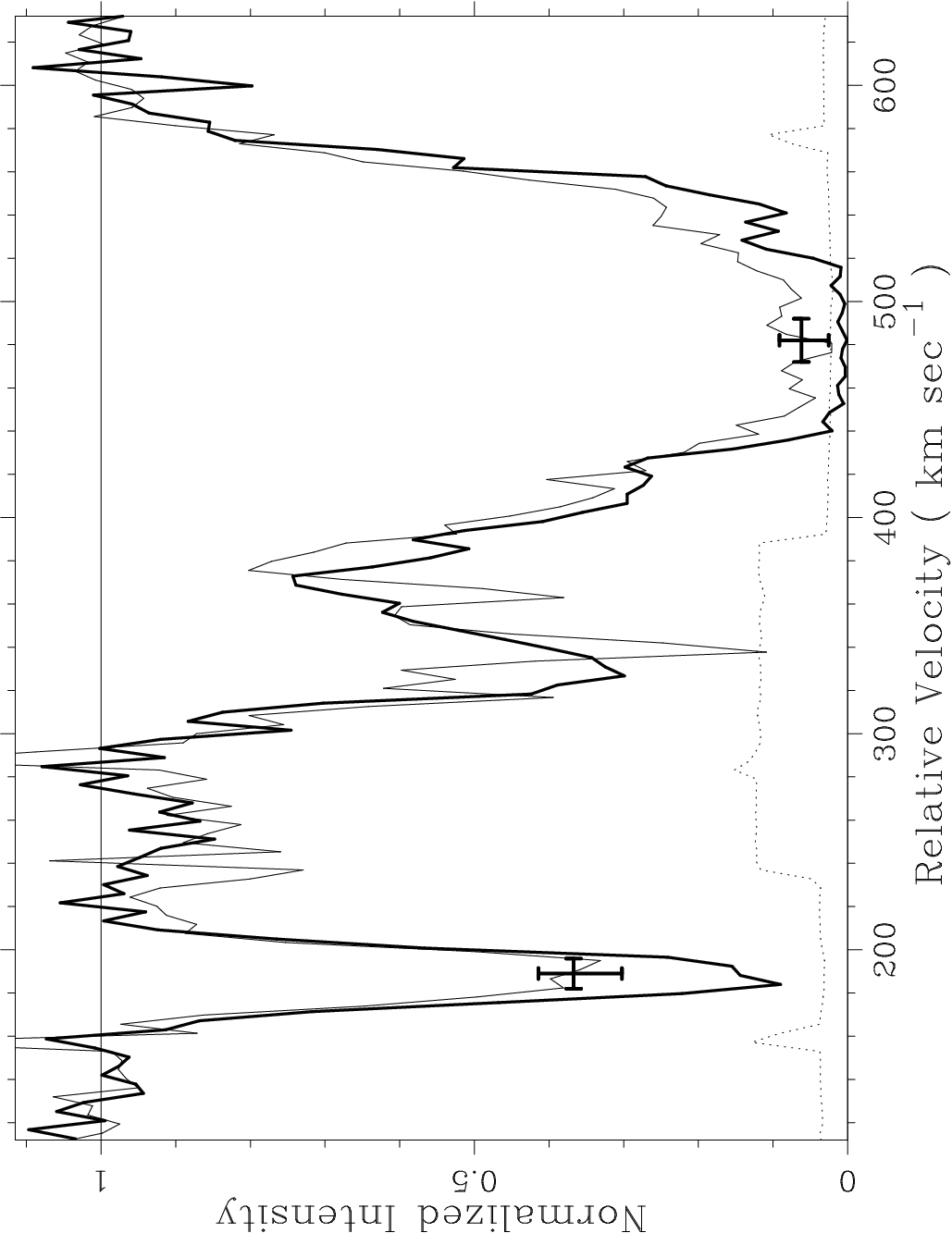]{The C IV associated absorption complex.  C IV
$\lambda$1548 is shown as a thin line and C IV $\lambda$1550 as a thick
line.  The optical depth of the 1550 \AA\ line has been scaled by a factor of
two.  Each bin is 4.2 km s$^{-1}$ wide and the 1$\sigma$ error per bin for
the 1548 \AA\ line is shown as a dotted line.  Also marked are the estimated
values of $1-C_f$ (with 3$\sigma$ error bars) for the lines at 190 km
s$^{-1}$ and 480 km s$^{-1}$ inflow relative to $z_e$.  \label{fig3}}

\figcaption[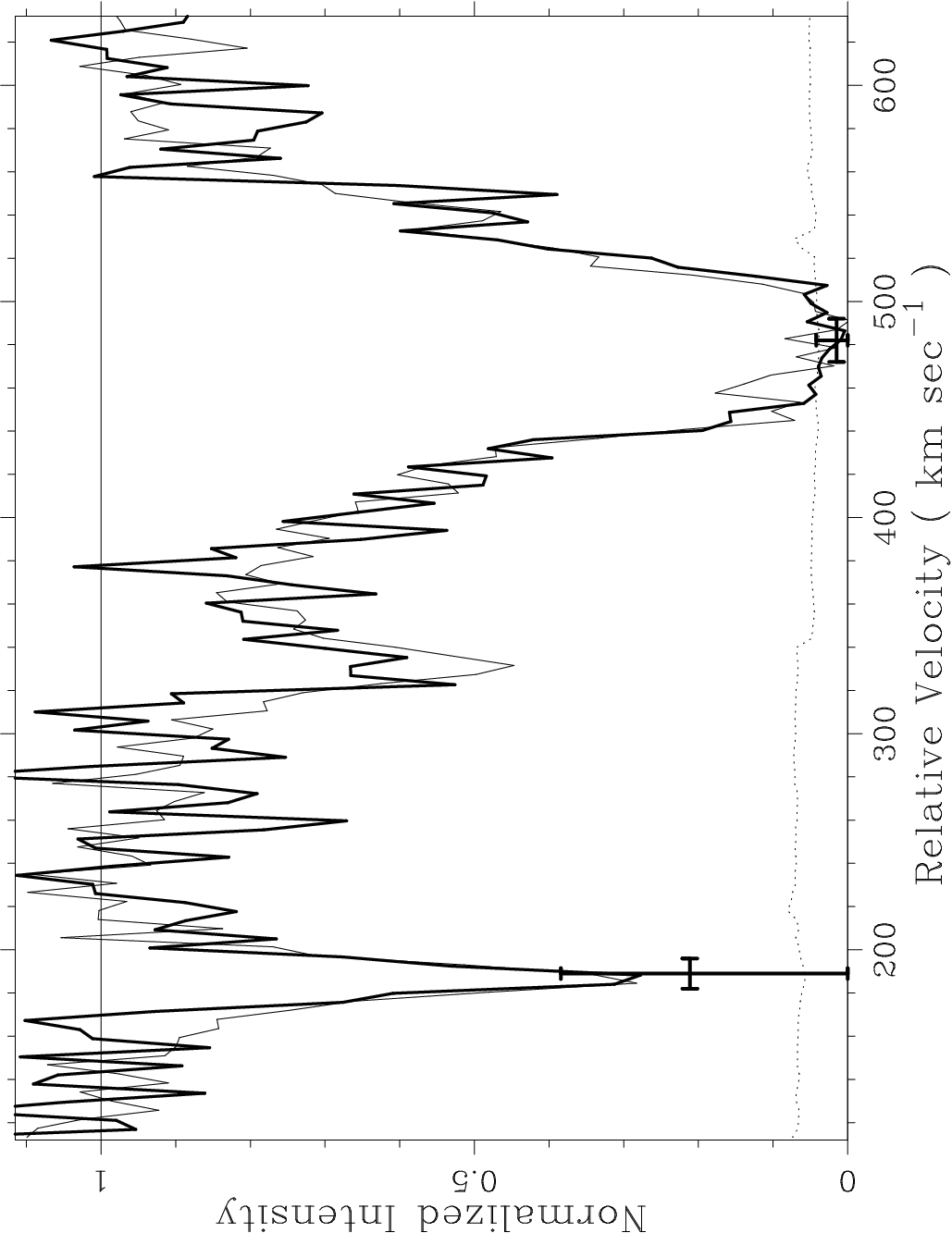]{The N V associated absorption complex.  N V
$\lambda$1238 is shown as a thin line and N V $\lambda$1242 as a thick line.
The optical depth of the 1242 \AA\ line has been scaled by a factor of two.
Each bin is 4.2 km s$^{-1}$ wide and the 1$\sigma$ error per bin for the 1238
\AA\ line is shown as a dotted line.  Also marked are the estimated values of
$1-C_f$ (with 1$\sigma$ error bars) for the lines at 190 km s$^{-1}$ and 480
km s$^{-1}$.  \label{fig4}}

\figcaption[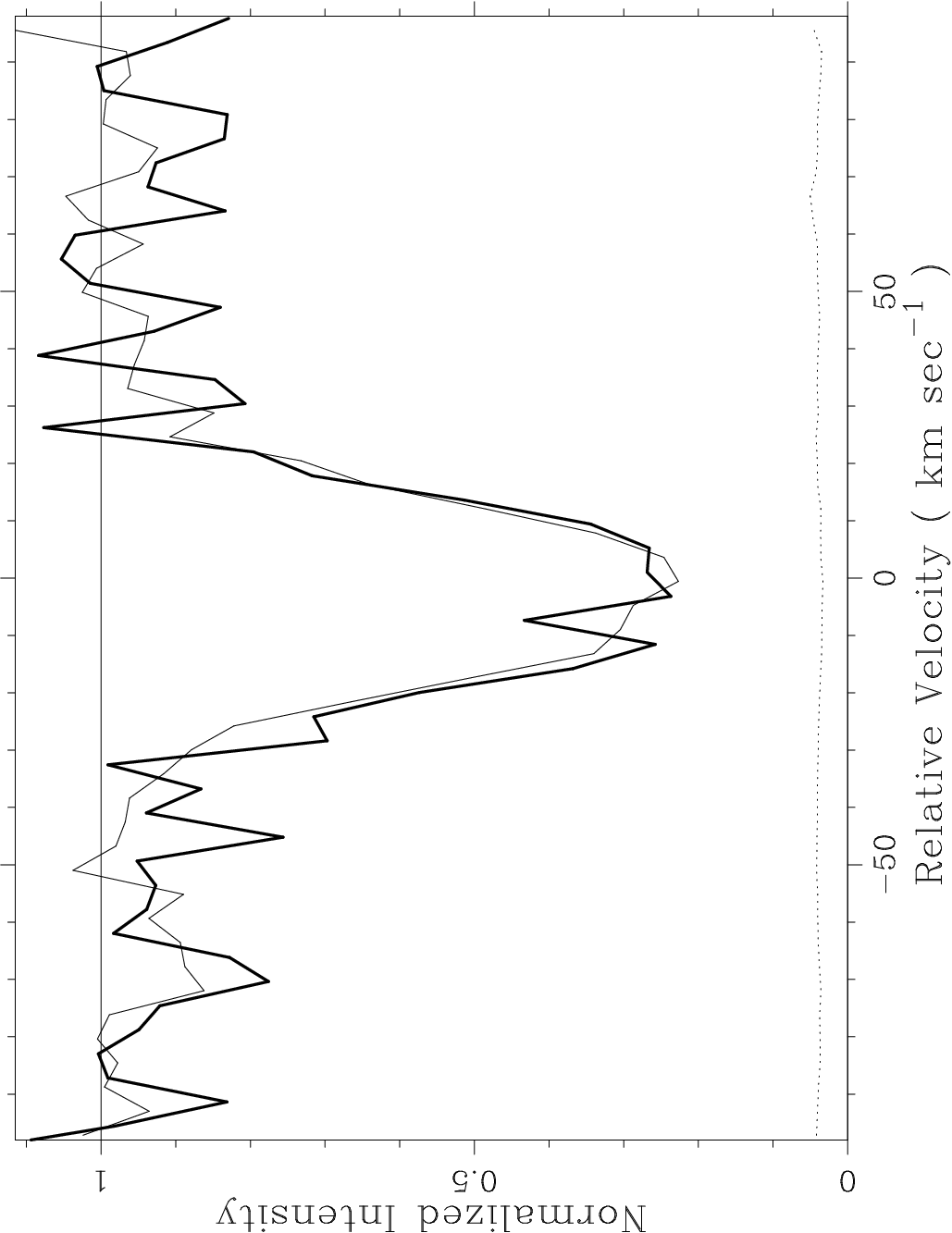]{ The C IV lines for a $z_a < z_e$ absorption
system at $z_a$=2.301 plotted as in figure 3 with the 1550 \AA\
optical depth increased by a factor of two.  Zero velocity is set to
$z_a$=2.301. \label{fig5}}

\clearpage
\begin{deluxetable}{lcccccc}
\tablenum{1}
\tablewidth{0pc}
\tablecaption{Absorption Line Measurements \label{tbl-1}}
\tablehead{
\colhead{Line}          & \colhead{$\lambda_{\rm obs}$} &
\colhead{Velocity\tablenotemark{a}}      & \colhead{$R$\tablenotemark{b}}  &
\colhead{$C_f$\tablenotemark{c}}         & \colhead{log(N)}              &
\colhead{log(N)} \\
\colhead{}              & \colhead{(\AA)}               &
\colhead{(km s$^{-1}$)} & \colhead{}                    &
\colhead{}              & \colhead{($C_f=1$)}           &
\colhead{($C_f<1$)}
}
\startdata
Ly$\alpha$ $\lambda$1215 &4083.0&+480& 0.092 & $>$0.91               & 14.7 & $>14.9$ \nl
\ion{N}{5} $\lambda$1238 &4160.7&+480& 0.033 & 0.98 $^{1.00}_{0.96}$ & 14.8 & $>14.8$ \nl
\ion{N}{5} $\lambda$1242 &4174.1&+480& 0.146 & \nodata               & 14.8 & $>14.8$ \nl
\ion{Si}{4} $\lambda$1393&4681.1&+480& 0.107 & 0.93 $^{1.00}_{0.88}$ & 13.6 & $>13.7$ \nl
\ion{Si}{4} $\lambda$1402&4711.4&+480& 0.254 & \nodata               & 13.6 & $>13.7$ \nl
\ion{C}{4} $\lambda$1548 &5199.8&+480& 0.063 & 0.94 $^{0.98}_{0.91}$ & 14.6 & $>14.8$ \nl
\ion{C}{4} $\lambda$1550 &5208.5&+480& 0.082 & \nodata               & 14.8 & $>15.0$ \nl
                         &      &    &       &      &      &      \nl
Ly$\alpha$ $\lambda$1215 &4086.9&+190& 0.154 & $>$0.85               & 14.0 & $>14.1$ \nl
\ion{N}{5} $\lambda$1238 &4164.8&+190& 0.426 & 0.79 $^{1.00}_{0.62}$ & 13.7 & $>13.8$ \nl
\ion{N}{5} $\lambda$1242 &4178.2&+190& 0.623 & \nodata               & 13.6 & $>13.7$ \nl
\ion{Si}{4} $\lambda$1393&4685.6&+190&\nodata& \nodata               &\nodata&\nodata \nl
\ion{Si}{4} $\lambda$1402&4715.9&+190&\nodata& \nodata               &\nodata&\nodata \nl
\ion{C}{4} $\lambda$1548 &5204.9&+190& 0.367 & 0.63 $^{0.70}_{0.59}$ & 13.6 & $>14.0$ \nl
\ion{C}{4} $\lambda$1550 &5213.5&+190& 0.372 & \nodata               & 13.8 & $>14.2$ \nl
\tablenotetext{a}{Relative to $z_e$=2.364.}
\tablenotetext{b}{Residual intensity at lowest point of absorption line.}
\tablenotetext{c}{Coverage fraction limits are 3$\sigma$ 
for \ion{C}{4}, 1$\sigma$ for \ion{N}{5} and \ion{Si}{4}.}
\enddata
\end{deluxetable}
\clearpage

\begin{figure}
\figurenum{1}
   \plotfiddle{Barlow.fig1.ps}{20.cm}{0.0}{97}{97}{-170}{-80}
\vskip 40pt
\caption{ see figure caption page}
\end{figure}
\clearpage
\begin{figure}
\figurenum{2}
   \plotfiddle{Barlow.fig2.ps}{20.cm}{0.0}{95}{95}{-170}{-100}
\vskip 40pt
\caption{ see figure caption page}
\end{figure}
\clearpage
\begin{figure}
\figurenum{3}
   \plotfiddle{Barlow.fig3.ps}{20.cm}{0.0}{95}{95}{-170}{-100}
\vskip 40pt
\caption{ see figure caption page}
\end{figure}
\clearpage
\begin{figure}
\figurenum{4}
   \plotfiddle{Barlow.fig4.ps}{20.cm}{0.0}{95}{95}{-170}{-100}
\vskip 40pt
\caption{ see figure caption page}
\end{figure}
\clearpage
\begin{figure}
\figurenum{5}
   \plotfiddle{Barlow.fig5.ps}{20.cm}{0.0}{95}{95}{-170}{-100}
\vskip 40pt
\caption{ see figure caption page}
\end{figure}


\clearpage

\begin{thebibliography}{}

\bibitem[Anderson et al.\ 1987]{and87}
    Anderson, S. F., Weymann, R. J., Foltz, C. B., 
    and Chaffee, F. H., Jr. 1987, \aj, 94, 278

\bibitem[Bahcall 1968]{bah68}
    Bahcall, J. N. 1968, \apj, 153, 679

\bibitem[Barlow et al.\ 1992]{bar92}
    Barlow, T. A., Junkkarinen, V. T., Burbidge, E. M.,
    Weymann, R. J., Morris, S. L., and Korista, K. T. 1992, \apj, 397, 81

\bibitem[Barlow 1993]{bar93}
    Barlow, T. A. 1993,
    Ph.D. thesis, Univ. of California, San Diego

\bibitem[Barlow 1995]{bar95}
    Barlow, T. A. 1995, \baas, 27, 872

\bibitem[Barthel 1988]{bar88}
    Barthel, P.D. 1988, in ESO Mini-Workshop on Quasar Absorption Lines,
    ed. P.A. Shaver, E.J. Wampler, and A.M. Wolfe (Garching: ESO), 79

\bibitem[Briggs et al.\ 1984]{bri84}
    Briggs, F. H., Turnshek, D. A., and Wolfe, A. M. 1984, \apj, 287, 549

\bibitem[Brotherton et al.\ 1994a]{bro94a}
    Brotherton, M. S., Wills, B. J., Steidel, C. C., 
    and Sargent, W.L.W. 1994a, \apj, 423, 131

\bibitem[Brotherton et al.\ 1994b]{bro94b}
    Brotherton, M. S., Wills, B. J., Francis, P. J., 
    and Steidel, C. C. 1994b, \apj, 430, 495

\bibitem[Carswell et al.\ 1976]{car76}
    Carswell, R. F., Coleman, G., Strittmatter, P. A., and 
    Williams, R. E. 1976, \aap, 53, 275

\bibitem[Cohen et al.\ 1995]{coh95}
    Cohen, M. H., Ogle, P. M., Tran, H. D., Vermeulen, R. C.,
    Miller, J. S., Goodrich, R. W., and Martel, A. R. 1995, \apj, 448, L77

\bibitem[Foltz et al.\ 1986]{fol86}
    Foltz, C. B., Weymann, R. J., Peterson, B. M., Sun, L., Malkan, M. A., and
    Chaffee, F. H., Jr. 1986, \apj, 307, 504

\bibitem[Foltz et al.\ 1988]{fol88}
    Foltz, C. B., Chaffee, F. H., Jr., Weymann, R. J., and 
    Anderson, S. F. 1988,
    in {\it QSO Absorption Lines: Probing the Universe,} 
    eds. J. C. Blades, D. A. Turnshek, and C. A. Norman
    (Cambridge: Cambridge Univ. Press), p. 53

\bibitem[Goodrich and Miller 1995]{goo95}
    Goodrich, R. W., and Miller, J. S. 1995, \apj, 448, L73

\bibitem[Hamann et al.\ 1995]{ham95}
    Hamann, F., Barlow, T. A., Beaver, E. A., Burbidge, E. M., Cohen, R. D.,
    Junkkarinen, V., and Lyons, R. 1995, \apj, 443, 606

\bibitem[Hamann 1996]{ham96}
    Hamann, F. 1996, submitted to \apjs

\bibitem[Hamann et al.\ 1996]{hamet96}
    Hamann, F., Barlow, T. A., Beaver, E. A., Burbidge, E. M., Cohen, R. D.,
    Junkkarinen, V., and Lyons, R. 1996, submitted to \apj

\bibitem[Hamann et al.\ 1996b]{hamet96b}
    Hamann, F., Barlow, T. A., and Junkkarinen, V. 1996,
    submitted to \apj

\bibitem[Horne 1986]{hor86}
    Horne, K. 1986, \pasp, 98, 609

\bibitem[Junkkarinen et al.\ 1995]{jun95}
    Junkkarinen, V. T., Beaver, E. A., Burbidge, E. M., Cohen, R. D., 
    Hamann, R. W., Lyons, R. W., and Barlow, T. A. 1995, \baas, 27, 872

\bibitem[Korista et al.\ 1996]{kor96}
    Korista, K. T., Hamann, F., Ferguson, J., and Ferland, G. 1996,
    \apj, 461, 641

\bibitem[Morris et al. 1986]{mor86}
    Morris, S. L., Weymann, R. J., Foltz, C. B.,
    Turnshek, D. A., Shectman, S., Price, C., 
    and Boroson, T. A., \apj, 310, 40

\bibitem[van Ojik et al. 1996]{oji96}
    van Ojik, R., R\"{o}ttgering, H. J. A., Miley, G. K.,
    and Hunstead, R. W. 1996, preprint.

\bibitem[Osterbrock 1993]{ost93}
    Osterbrock, D. E. 1993, \apj, 404, 551

\bibitem[Petitjean et al.\ 1994]{pet94}
    Petitjean, P., Rauch, M., and Carswell, R. F. 1994, \aap, 291, 29

\bibitem[Sargent et al.\ 1988]{sar88}
    Sargent, W. L. W., Boksenberg, A., and Steidel, C. C. 1988, \apjs, 68, 539

\bibitem[Savage and Sembach 1991]{sav91}
    Savage, B. D. and Sembach, K. R. 1991, \apj, 379, 245

\bibitem[Schmidt and Olsen 1968]{sch68}
    Schmidt, M., and Olsen, E. T. 1968, \aj, 73, S117

\bibitem[Tytler and Fan 1992]{tyt92}
    Tytler, D., and Fan, X. M. 1992, \apjs, 79, 1

\bibitem[Wampler et al.\ 1993]{wam93}
    Wampler, E. J., Bergeron, J., and Petitjean, P. 1993, \aap, 273, 15

\bibitem[Wampler et al.\ 1995]{wam95}
    Wampler, E. J., Chugai, N. N., and Petitjean, P. 1995, \apj, 443, 586

\bibitem[Weymann et al.\ 1979]{wey79}
    Weymann, R. J., Williams, R. E., Peterson, B. M., and
    Turnshek, D. A. 1979, \apj, 234, 33

\bibitem[Williams et al.\ 1975]{wil75}
    Williams, R. E., Strittmatter, P. A., Carswell, R. F., 
    and Craine, E. R. 1975, \apj, 202, 296

\bibitem[Wills et al.\ 1993]{wil93}
    Wills, B. J., Brotherton, M. S., Fang, D., Steidel, C. C., and
    Sargent, W. L. W. 1993, \apj, 415, 563

\end{thebibliography}
\end{document}